\documentclass[aps,pra,twocolumn,superscriptaddress,amsmath,amssymb]{revtex4}
\usepackage{graphicx}
\usepackage{epsfig}
\usepackage{bm}
\usepackage{multirow,amssymb,amsbsy,amsmath}

\newcommand*{\ket}[1]{\mathopen{|}#1\mathclose{\rangle}}
\newcommand*{\bra}[1]{\mathopen{\langle}#1\mathclose{|}}

\begin{document}
\title{Scaling of entanglement between separated blocks in spin chains at criticality}
\author{H. Wichterich}
\affiliation{
    Dept.~of Physics and Astronomy, %
    University College London, %
    Gower Street, %
    WC1E 6BT London, %
    United Kingdom
}
\author{J. Molina-Vilaplana}
\affiliation{
    Dept.~of Systems Engineering and Automation,
    Technical University of Cartagena,
    Campus Muralla del Mar s/n.,
    30.202 Cartagena,
    Murcia,
    Spain
}
\author{S. Bose}
\affiliation{
    Dept.~of Physics and Astronomy, %
    University College London, %
    Gower Street, %
    WC1E 6BT London, %
    United Kingdom
}

\begin{abstract}
We compute the entanglement between separated blocks in certain spin
models showing that at criticality this entanglement is a function of the ratio of the
separation to the length of the blocks and can be written as a product of a power law
and an exponential decay. It thereby interpolates between the entanglement
of individual spins and blocks of spins. It captures features of
correlation functions at criticality as well as the monogamous
nature of entanglement. We exemplify invariant features of this
entanglement to microscopic changes within the same universality
class. We find this entanglement to be invariant with respect to
simultaneous scale transformations of the separation and the length
of the blocks. As a corollary, this study estimates the entanglement between
separated regions of those quantum fields to which the considered
spin models map at criticality.
\end{abstract}
\maketitle

\begin{figure}
 \epsfig{figure=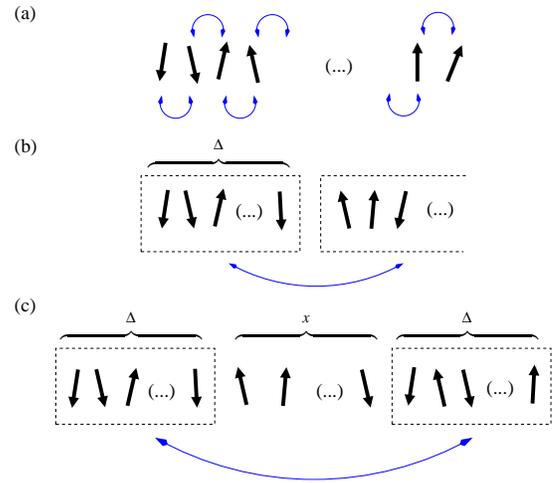, width=0.4\textwidth}
\caption{Schematic of different types of entanglement at a quantum critical point. Panel (a): Critical behavior of entanglement of pairs of
spins, exhibiting very limited range. Panel (b): Entanglement entropy of two contiguous and complementary blocks of spins, obeying a universal
scaling law at criticality. Panel (c): Entanglement between distant blocks of spins as a function of their size $\Delta$ and separation $x$ in a
spin chain of finite but large length $N=2\Delta+x$~. \label{fig.pic}}
\end{figure}

Correlations have long been a central object of study in condensed matter with attention recently drawn to entanglement -- unique correlations
possible only in quantum mechanics. The presence of entanglement inside condensed matter systems \cite{Osterloh2008,Vedral2008} is also
experimentally supported \cite{Ghosh2003}. These {\it quantum} correlations become particularly interesting at quantum phase transitions (QPT),
which occur at zero temperature as the relative strengths of interactions in a many-body system are varied \cite{Sachdev}. At a QPT, in general,
the entanglement between individual spins is non-zero only for very small separations between the spins \cite{Osterloh2002} (Fig.~\ref{fig.pic},
panel (a)). On the other hand, the entanglement between adjacent blocks of spins (which cannot by definition, have a separation) diverges with
the length of the blocks \cite{Vidal2003,Latorre2004,Jin2004} (Fig.~\ref{fig.pic}, panel (b)).
However, an intermediate situation where one considers the entanglement {\em between} blocks of spins which are separated (i.e.,
non-complementary) is an open problem. It would be interesting to study how this entanglement scales, i.e., varies with the size of the blocks
$\Delta$ and their separation $x$, at a QPT. Here, we conduct such a study (see, Fig.\ref{fig.pic}, panel (c)) and find that both the short
ranged nature of spin-spin entanglement and divergent nature of adjacent block entanglement can be recovered qualitatively as limiting cases of
the expression for the entanglement of non-complementary blocks. This can be viewed as an interpolation between known behaviors of entanglement
in the aforementioned limits. Though, what we compute is a form of bipartite entanglement, it also captures aspects of the multipartite
entanglement in the system, as we will discuss later.

Note that two other information theoretic quantities involving non-complementary blocks of a quantum many body system have recently drawn much
attention. One of them, the von Neumann entropy of disjoint blocks quantifies the entanglement of these blocks with respect to their complement,
but not that {\em between} them \cite{Facchi2008,Calabrese2009}. The second, mutual information, does quantify the correlations between the
blocks, but cannot be considered as a measure of entanglement (i.e., the ``quantum" part of the correlations) \cite{Furukawa2009,Calabrese2009}.
Seperated blocks will, in general, be in a mixed state. Unfortunately, this inhibits the use of von Neumann entropy, for which analytical
methods exist \cite{Jin2004}, to quantifiy their entanglement and more involved measures\cite{Vidal2002} are necessary (as acknowledged in
Ref.~\cite{Calabrese2009}). Unfortunately, even for solvable free fermionic models there are, in general, no known analytic methods to compute
these more involved measures. Thus, we choose a numerical approach which is particularly suited to our goal (to be justified later).

While our principal result is the functional form of the entanglement of non-complementary blocks, after computing the entanglement we notice as
an aside, that it depends only on the ratio $\mu\equiv\frac{x}{\Delta}$ and therefore exhibits ``scale invariance". In other words, at the
considered critical points the block-block entanglement is invariant to simultaneous scale transformations $x\rightarrow bx,\, \Delta\rightarrow
b\Delta$. This is commensurate with the mapping of the citical models to a conformal field theory (CFT) in the thermodynamic limit. This is
still an interesting observation, because we never explicitly use results from CFT, but base our study on finite but large spin chains. Previous
computations of distant block entanglement have been for linear chains of quantal harmonic oscillators \cite{Audenaert2002}. Entanglement
between noncomplementary regions of size $R$ of quantum fields have previously been found to decay no faster than $\sim \mathrm{exp}[-(L/R)^2]$
(for Dirac fields \cite{Silman2007}) and $\sim \mathrm{exp}[-(L/R)^3]$ (for scalar fields \cite{Reznik2005}) where $L$ is the distance between
the regions.

We focus on the finite 1D XY model, consisting of a number $N$ of spins $\frac{1}{2}$ arranged on a regular lattice with nearest neighbour
interactions and subject to an external magnetic field. The Hamiltonian of the spin chain reads
\begin{align}\label{eq.hamiltonian}
 H=-\sum\limits_{k=1}^{N-1}\left(\frac{1+\gamma}{2}\,\sigma_k^{\text x}\sigma_{k+1}^{\text x}+\frac{1-\gamma}{2}\,\sigma_k^{\text y}\sigma_{k+1}^{\text y}\right)-\sum\limits_{k=1}^{N}\lambda\sigma^{\text z}_k~.
\end{align}
Here, $\sigma^{a}$ $(a = x,y,z)$ denote the Pauli matrices and we assume open boundary conditions. This model \cite{Latorre2004} features a QPT
at $\lambda=\lambda_{\text c} =1$ for any anisotropy $\gamma$. The XX model, $\gamma=0$, is also critical for field values $\lambda\in[0,1]$.
The critical regions of the model can be categorized into different universality classes: Ising and XX universality class for
$(\gamma\in(0,1],\lambda=1)$ and $(\gamma=0,\lambda\in[0,1])$ respectively \cite{Latorre2004}. In view of these properties, we choose three
critical points: (i) the critical Ising $(\gamma=1,\lambda=1)$, (ii) a critical XY $(\gamma=0.5,\lambda=1)$ and (iii) the critical XX at
vanishing field $(\gamma=0,\lambda=0)$. Our results show that for (i)-(iii) distant entanglement of blocks is scale invariant, and further (i)
and (ii) show a common universal behavior with respect to this entanglement measure.

Our approach is based on the Density Matrix Renormalisation Group
(DMRG) technique \cite{White1992}. We extend this method to extract from the ground state wave function $\ket{\psi}$
the reduced density operators $\rho_{SE}=\mathrm{Tr}_{\overline{SE}}\ket{\psi}\bra{\psi}$ of system (S) and
environment (E) blocks, both of equal length $\Delta$ and separated
by a number $x$ of consecutive spins (see Fig.\ref{fig.pic}, panel
(c)), as will be outlined in detail in the Appendix. The DMRG
procedure optimises the decimated set of $M$ basis states per block
in such a way, that the maximum entanglement entropy $\mathcal{S}$
of
all possible bipartitions (which arise from cutting the chain at a particular bond $l$ into left and right part) is retained, 
rendering the study of critical phenomena particularly delicate, due to the fact that here $\mathcal{S}$ diverges with block size \cite{Vidal2003,Latorre2004,Jin2004}. 
In Ref. \cite{Tagliacozzo2008} it was found that even at criticality DMRG faithfully recovers prototypical features of critical ground states, such as, e.g., polynomial decay of correlators, as long as $M$ is chosen sufficiently large.
\begin{figure}
 \epsfig{figure=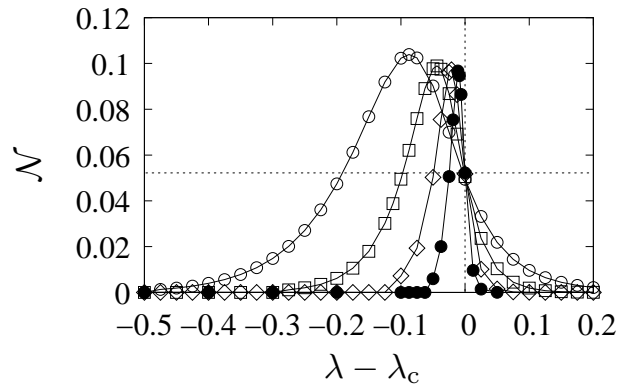, width=0.45\textwidth}
\caption{Scaling of distant block entanglement with system size at and in the vicinity of the critical point $\lambda-\lambda_{\text c}=0$ of the transverse Ising model ($\gamma=1$ in Eq.~\ref{eq.hamiltonian}). The data correspond to a fixed ratio $\mu=\frac23$, and different system sizes $N=32$ (open circles), $N=64$ (squares), $N=128$ (diamonds) and $N=256$ (filled circles). The crossing of dashed lines highlights the converged, scale invariant point for $N \gtrsim 128$, which is $(\lambda,\,\mathcal{N})\cong(\lambda_c,\,0.052) $ for this ratio. \label{fig.scan}}
\end{figure}
 Note that DMRG faithfully reproduces the {\it actual} entanglement between the two separated blocks. We are
 {\em neither} presuming a scale invariant behaviour of the entanglement {\em a priori} \cite{Vidal2007a} to compute the representation
 of the critical ground state, {\em nor} are we course-graining
 blocks of spins into effective spins \cite{Kargarian2008}.
 Thus, the fact that entanglement is only a function of $\mu$ is an outcome of our calculations rather than a prior input.

The reduced density operators $\rho_{SE}$ carry the information on
the entanglement between the blocks $S$ and $E$. As $\rho_{SE}$ is a
mixed state, the block entropy is inappropriate as a
measure of the entanglement. We thus have to use the negativity
$\mathcal{N}\equiv(\sum_i \vert a_i\vert - 1)$
 where $\vert a_i\vert$ denote the modulus of the eigenvalues
of the partial transpose $(\rho_{SE})^{\text{T}_S}$ of $\rho_{SE}$ with respect to the subsystem $S$, i.e., $\bra{w^S_i\,
w^E_j}\rho_{SE}^{T_S}\ket{w^S_k\, w^E_l}=\bra{w^S_k\, w^E_j}\rho_{SE}\ket{w^S_i\, w^E_l}$ \cite{Vidal2002,Plenio2005}. $\lbrace
\ket{w^S}\rbrace$ and $\lbrace \ket{w^E}\rbrace$ are the orthogonal basis states of $S$ and $E$ respectively, chosen by the DMRG procedure. This
is a widely used genuine measure of quantum correlations (entanglement monotone\cite{Plenio2005}) and provides a bound to the fidelity of
teleportation with a single copy of the state\cite{Vidal2002}. $\mathcal{N}$ depends on the size of the entangled regions $\Delta$ and their
separation $x$. Once $x$ exceeds the range of pairwise entanglement of spins, an individual spin in $S$ is not entangled with an individual spin
in $E$. Nonzero $\mathcal{N}(\rho_{SE})$ will then genuinely signal multipartite entanglement\cite{Osterloh2008,Patane2007}. We find that at the
critical regions of the XY model, Eq.\ref{eq.hamiltonian},
$\mathcal{N}$ is a function of the
ratio $\mu\equiv\frac{x}{\Delta}$ {\em only}
and shows universal behaviour.
We focus on those values of $\mu\geq 0.1$, to avoid corrections stemming from finite $N$, which enforce a finite value on $\mathcal{N}$ (in the limit of adjacent blocks), whereas it is
expected to diverge for infinite $N$. We also restrict to $\mu\leq 3$, because beyond that $\mathcal{N}$ assumes an order of magnitude comparable to the accuracy of the numerics.
We checked the convergence of $\mathcal{N}$ with the number of kept basis states
per block $M$, and are able to produce accurate data reliable at
least up to the 5th (3rd) decimal place for calculations in models
of the Ising (XX) universality class.

Naturally, one might ask why we use a numerical method when the considered models are exactly solvable by a transformation to free fermions,
with the ground states being fully specified by second moments. In the case of their bosonic counterparts powerful methods have been established
\cite{Audenaert2002} to compute the negativity of separated blocks in terms of the second moments, yielding numerically exact results. However,
this method does not readily generalize to fermions. Thus with current knowledge, the only option would be to explicitly construct $\rho_{SE}$
in the standard (spin) basis. Even after exactly knowing the second moments it is practically unfeasible to write down (store for computation)
$\rho_{SE}$ for large chains. Hence, we exploit the convenient and effective representation $\lbrace\ket{w^S}\otimes\ket{w^E}\rbrace$ of
$\rho_{SE}$ arising in DMRG.

\begin{figure}
 \epsfig{figure=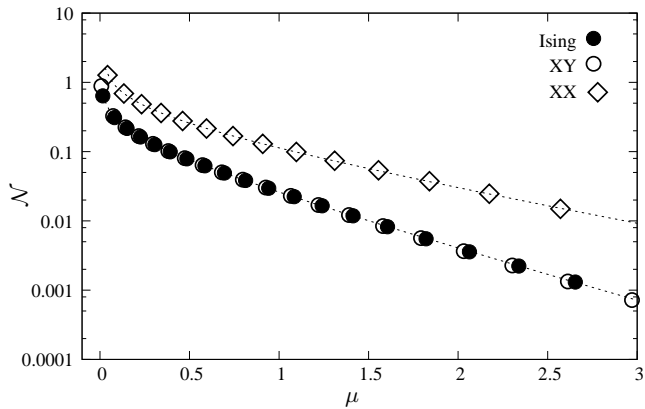, width=0.47\textwidth}
\caption{Negativity $\mathcal{N}$ as a function of the ratio $\mu=\frac{x}{\Delta}$, for the critical XX at vanishing field (diamonds), critical
XY (open circles) and critical Ising (filled circles) models. The data for the critical XX model are fitted according to our ansatz, revealing
the parameters $h=0.47,\,\alpha=0.96$ (dashed line). The critical Ising model is fitted accordingly, with parameters $h=0.38,\,\alpha=1.68$. All
data correspond to selected subsets of possible ratios $\mu$, in favour of visibility. \label{fig.neg}}
\end{figure}
 Turning to our results shown in Fig.~\ref{fig.scan}, we see that for
a fixed ratio $\mu$ the negativity $\mathcal{N}(\rho_{SE})$ shows
the distinctive feature of scale invariance at the critical point
$\lambda=\lambda_{\text c}=1$ of the transverse Ising model. A
similar behaviour can be observed for the other critical models
considered here and for other ratios $\mu$.

 Most important and interesting is to enquire how $\mathcal{N}(\rho_{SE})$ depends on
different ratios $\mu$ in the scale free point (once data have properly converged with system size $N$). From Fig.\ref{fig.neg} we see that for
$\mu\lesssim 2.5$ the data for the two models $\gamma=1$ and $\gamma=0.5$ match near perfectly signalling {\em universality}. One further
recognises a {\em generic shape} of $\mathcal{N}(\rho_{SE})$ as a function of $\mu$ for all three models under consideration, its decay
exhibiting a polynomial onset and an exponential tail for $\mu\gtrsim 0.2$. Given the polynomial decay of the correlation functions with spatial
separation, it seems plausible that quantum correlations will inherit signatures thereof. However unlike correlation functions, entanglement is
also {\em monogamous} \cite{Coffman2000}: the entanglement of one block with another can only be a fraction of its total entanglement with the
rest of the chain. This is another way in which the bipartite entanglement between blocks captures the multipartite nature of entanglement in a
system (i.e. between the blocks in a three block system). At the limit of individual spins this leads the concurrence to decay very fast with
separation. Generalizing this trend, one may expect an exponential decay of the entanglement of blocks with separation. In view of these
considerations, we make the ansatz
\begin{equation}
\mathcal{N}(\rho_{SE})\sim \mu^{-h}\, e^{-\alpha\mu}
\end{equation}
with real parameters $h$ and $\alpha$. This ansatz is vindicated by providing an excellent fit for our data, though the precise values for
$\alpha$ and $h$ inferred from the fitting are somewhat sensitive (in the second decimal place) to the chosen fitting interval. One of the
intriguing open questions is how the exponents $\alpha$ and $h$ are related to known critical exponents. Our fit suggests, in the case of the
XX-model, $\alpha=0.96,\,h=0.47$ and for the XY models $\alpha=1.68,\, h=0.38$.
Instead, if one fitted the logarithmic negativity\cite{Vidal2002}, in the present notation defined by
$E_{LN}\equiv\log_2(\mathcal{N}+1)$ , to the same ansatz
this would lead to $h=0.33 \sim 1/3$. The same number was also numerically observed\cite{Marcovitch2008} for finite blocks
in infinite harmonic oscillator chains. This is further confirmation of the correctness of our work as both models map
to the same massless bosonic field theory in the continuum limit.
Note that in the limit of vanishing separation to block length ratio (vanishing $\mu$) $\mathcal{N}(\rho_{SE})\sim\mu^{-h}$ and thus diverges as
$\Delta^h$ for fixed separation (the logarithmic negativity diverges as $h\, \mathrm{log}_2\,\Delta$). Thereby qualitatively it reproduces the
features of block entropy (similar divergent behaviour was reported\cite{Orus2008} for other measures, also including multipartite
settings\cite{Wei2005}), though the negativity and block entropy are not related by any known simple formula in general -- thus one would not
exactly coincide with the other in any limit. In the limit of large $\mu$, i.e., small blocks of very distant spins the exponential part will
severely dominate and ensure that their entanglement is nearly zero. This is expected because of the limited entangling capacity of small
blocks. This capacity is exhausted by being entangled to their close neighbors which are granted a larger share of the entanglement because of
the nearest neighbor nature of interactions.

Summarizing, we have investigated the entanglement between separated blocks of spins at critical points of spin chains.
We have conjectured an ansatz for the functional form of the scaling of this entanglement with separation and size of the blocks and shown it to
be an excellent fit for our data. This functional form involves only a ratio of length scales and therefore exhibits an interesting scale
invariance, as may be expected for models which can be mapped to a scale free CFT. It further qualitatively encompasses two known limits of
entanglement of adjacent blocks and that of pairs of spins. We further exemplified invariant features of this entanglement to microscopic
changes within the same universality class. Interesting open questions are relating the numerically inferred coefficients $h$ and $\alpha$ to
known critical exponents, and constructing measurable variables defined on the separated blocks which will witness their entanglement.
\appendix*
\section{Extracting $\rho_{SE}$ within DMRG}
The canonical DMRG representation of the ground state $\ket{\psi}$ is given by the schematic {\bf S} $\bullet\,\bullet$ {\bf E}, where the
boldface fonts designate a decimated representation of system and environment block respectively, which may contain a large number of physical
{\it sites}. The bold dots represent two neighboring sites, represented in the basis
$\lbrace\,\ket{\uparrow,\uparrow},\,\ket{\uparrow,\downarrow},\,\ket{\downarrow,\uparrow},\,\ket{\downarrow,\downarrow}\,\rbrace$. The location
of this pair, indexed by the connecting bond $l\in [1,2,\cdots,N-1]$, may be shifted to an arbitrary position in the corresponding {\it real}
lattice. In particular for even $N$ we may choose $l$ such that {\bf S} and {\bf E} represent blocks of equal number $\Delta$ of spins, i.e. the
symmetric configuration. This is the departure point for our study of the entanglement of the blocks {\bf S} and {\bf E}, separated by a number
$x$ of lattice sites, i.e. initially $x=2$. In DMRG the representation of {\bf S} arises in the course of a growth procedure from the direct
product of a smaller system block {\bf S'} (in our case containing $\Delta-1$ spins) and a single site $\bullet$. The product {\bf S'}
$\otimes\,\bullet$ is then transformed into a decimated basis of dimension $M$, designated as ({\bf S'}\,$\bullet$). {\bf S'} is again a common
description of a yet smaller block and a single site. This results in a nested representation {\bf S} $\leftrightarrow$ (({\bf
S''}\,$\bullet$)\,$\bullet$) and so forth. In a similar fashion {\bf E} arises from such a nesting procedure, {\it i.e.} {\bf E}
$\leftrightarrow$ ($\bullet$\,($\bullet$\,{\bf E''})). As the basis transformations are stored, we may also  go back to a product
representation, {\it i.e.} ({\bf S'}$\,\bullet$) $\rightarrow$ {\bf S'} $\otimes\,\bullet$. This allows us to successively obtain the reduced
density operators of the separated blocks $\rho_{SE}$, $\rho_{S'E'}$, $\rho_{S''E''}$, and so on at a modest computational expense: The matrices
representing the various $\rho_{SE}$ are at most of dimension $M^2\times M^2$. The subsequent diagonalization of $(\rho_{SE})^{T_S}$ limits the
maximal $M$ we can allow, i.e., in our case $M\leq 60$, and conversely, the maximal $N$ that can be simulated for a given accuracy. A figure of
merit for the accuracy of the DMRG procedure is the truncated weight $\epsilon\equiv\sum_{i>M}w_i$ where $w_j\geq w_i\Leftrightarrow i>j$ stem
from the Schmidt decompostion $\ket{\psi}=\sum_{i} \sqrt{w_i} \ket{w_i^S}\otimes\ket{w_i^E}$. We require $\epsilon < 10^{-10}(10^{-8})$ which is
satisfied for chain lengths as large as $N=1024\,(96)$ in the anisotropic XY (XX) model, implying a similar accuracy for the block entropy
$\mathcal{S}$. With this $\epsilon$ the negativity has converged up to the 5th(3rd) decimal place.

 {\em Note added.} -- When finalizing this
paper we became aware of the independent work on entanglement of noncomplementary regions of a scalar massless field \cite{Marcovitch2008},
whose results match with our results of the XX model.

\begin{acknowledgments}
HW is supported by the EPSRC, UK. JMV acknowledges the Spanish Office for Science and Technology program ``Jose Castillejo'' and Fundacion
Seneca Murcia. SB acknowledges the EPSRC, UK, the EPSRC sponsored QIPIRC, the Royal Society and the Wolfson Foundation.
\end{acknowledgments}

\end{document}